\documentclass[12pt,a4paper]{article}
\usepackage{graphicx,amsmath,amsfonts}
\def \ba{\begin{eqnarray}}\def\ea{\end{eqnarray}}
\def\bc{\begin{center}}\def\ec{\end{center}}
\def\nn{\nonumber\\}
\title{\huge \bf  Final state interaction  in kaons decays.}
\author{\bf S.R.Gevorkyan \footnote{Corresponding author:
gevs@jinr.ru (S.Gevorkyan)} \footnote{On leave of absence from
Yerevan Physics Institute}, A.V.Tarasov,
O.O.Voskresenskaya\footnote{On leave of absence from Siberian
Physical Technical Institute}}
\date{}
\begin{document}
\maketitle
 \bc
Joint Institute for Nuclear Research, 141980 Dubna,
Russia \ec
\begin{abstract}
The kaons decays   to the  pairs of  charged and neutral pions  are  considered in the
 framework of the non-relativistic quantum mechanics. The general expressions  for
the decay amplitudes to the two different channels  accounting for the strong
 interaction between pions  are obtained. The developed approach   allows one to estimate
the contribution of terms of any order  in strong interaction  and correctly takes into account
 the electromagnetic interaction  between the  pions in the final state.
\end{abstract}
\section{Introduction}
It has long been known ~\cite{gribov58,gribov61}  that the  K-mesons decays with a
pions in the final state can give unique information on the pions s-wave scattering lengths $a_0,a_2$ ,
whose  values are predicted by Chiral Perturbation Theory (ChPT) with  high accuracy~\cite{colangelo01}.\\
Recently the high quality data on  $K^\pm\to \pi^\pm\pi^0\pi^0$ decays have
been obtained  by  NA48/2 collaboration  at CERN \cite{batley06}. The dependence
of the decay rate on the invariant mass of neutral pions $M^2=(p_1+p_2)^2$  reveals
a prominent anomaly (cusp) at the charged pions production  threshold  $M_c^2=4m_c^2$.\\
As was  explained  in ~\cite{cabibbo04,cabibbo05} this anomaly is due to
the possibility for the  kaon  to  decay  to three charged pions, which after
charge exchange reaction  $\pi^+\pi^-\to \pi^0\pi^0$  gives the observed neutral pions.
This possibility is provided by mass difference of charged and neutral pions.
The detail consideration of this decay using the technique of non-relativistic
 field theory~\cite{colangelo06} or  ChPT ~\cite{gamiz07} supports  the proposed picture.\\
 Nevertheless there are two  challenges crucial in scattering lengths extraction
 from  kaons decays. One needs  a reliable way  to estimate the contribution
 of higher order terms in strong interaction and calculates  the electromagnetic interaction among the charged pions.\\
 These  issues are very close connected with each other. Calculation of  the electromagnetic
 interaction  in every order of strong interaction~\cite{bisseger09} doesn't solve the problem of   bound states (pionium atoms), as  to take into account  electromagnetic interaction leading to unstable  bound states  one needs expressions  for decay amplitudes including the  strong interaction between pions in  all orders~\cite{gevorkyan07}.
The problem of correct accounting  of the electromagnetic effects are also necessary in a wide class of decays with
two pions in the final state as for instance  $K_{e4}$ decay~\cite{gevorkyan1,gevorkyan2}.\\
The phenomenon of cusp in elastic scattering at the threshold
relevant to inelastic channel is known for many years and was widely
discussed in the framework of non-relativistic quantum
mechanics~\cite{wigner48,breit57,baz57}.For the elastic process
$\pi^0\pi^0\to\pi^0\pi^0$ this anomaly at the $\pi^+\pi^-$ threshold
was firstly discussed in the framework of  ChPT in ~\cite{meisner97}.
In the present work  we consider the  kaon decay to  pion  pairs  with pions of different
masses. Using the well known results of quantum mechanics we obtain  the matrix elements
for decay $K\to \pi\pi$  where the final  pions  consist from pions of different  masses
($\pi^+\pi^-,\pi^0\pi^0$).

\section{Two channel  decay}
We  are interested  in  two channel decay of  kaon  to the  pion pairs in the final state,
where the pions  in the pair can be  neutral  or  charged . The well examples
are  $K_L\to \pi\pi$ as well as  $K^\pm\to\pi^+\pi^-e^\pm\mu$ ($ K_{e4}$ decay).
In what follows all quantities relevant to the neutral pions pair ($\pi^0\pi^0$) are labeled by
index "n", whereas  the charged pions pairs  ($\pi^+\pi^-$)  are  labeled by index  "c".\\
 We do not consider here the electromagnetic interaction in the pair, the effect  discussed  in our previous work~\cite{gevorkyan2}. Our main goal is to obtain the  matrix elements of the kaon decay to the pion
 pair  accounting for different masses of neutral and charged pions and the possibility  of charged exchange
 reaction $\pi^+\pi^-\to \pi^0\pi^0$  and  the elastic scattering of pions in the final state.\\
The general form of matrix element for kaon  decay  to the final state with  two charged  or neutral pions
 can be written in the operator form:
\ba
M_c=\int \Psi_c^+(r)M_0(r)d^3r;\hspace{1cm}
 M_n=\int \Psi^+_n(r)M_0(r)d^3r
\ea
The two component  operator   $M_0=\left (\begin{array}{c}M_c^{(0)}(r)\\M_n^{(0)}(r)\end{array}\right)$,
where $M_c^{(0)}(r),M_n^{(0)}(r)$ are the matrix elements of  kaon decay to noninteracting charged
and neutral pions pairs, while $\Psi_c(r), \Psi_n(r)$  are the  appropriate two component  wave  functions. \\
These wave functions would  satisfy to  couple  Shr\H{o}dinger equations \footnote{Throughout  this paper we   restricted   by    s-wave   $\pi\pi$   scattering in the final state.}
\ba
 -\Delta\Psi_c(r)+U_{cc}\Psi_c(r)+U_{cn}\Psi_n(r)&=&k_c^2\Psi_c(r)\nn
-\Delta\Psi_n(r)+U_{nn}\Psi_n(r)+U_{nc}\Psi_c(r)&=&k_n^2\Psi_n(r)
\ea
where  $U_{ij}$ are the strong potentials describing  elastic $cc\to cc;  nn\to nn$  scattering and
charge exchange reaction $cn\to cn$ . $k_c,k_n$ are the  charge and neutral pions  momenta in the appropriate center of mass system.\\
According to the general principles of  scattering theory    the asymptotic behavior of the  wave functions
$\Psi_c(r) ,\Psi_n(r)$  can be written through  the  s-wave amplitudes  $f_{cc}, f_{nn}, f_{cn}, f_{nc}$
in the following  form:
\ba
\Psi_c(r)&=&\left (\begin{array}{c}\frac{sin{k_cr}}{k_cr} \\ 0 \end{array}\right)+
\left (\begin{array}{c}\frac{e^{-ik_cr}}{r}f^*_{cc} \\ \frac{e^{-ik_nr}}{r}f^*_{nc} \end{array}\right)\nn
\Psi_n(r)&=&\left (\begin{array}{c}0\\ \frac{sin{k_nr}}{k_nr}  \end{array}\right)+
\left (\begin{array}{c}\frac{e^{-ik_cr}}{r}f^*_{cn} \\ \frac{e^{-ik_nr}}{r}f^*_{nn} \end{array}\right)
\ea
 The first columns in these expressions describe  the noninteracting s-waves pions pairs, whereas the second columns correspond to the  interacting charge and neutral  pions pair  in the far asymptotic of corresponding wave function.\\
 One can rewritten   these equations  through the elements  of  appropriate S-matrix  ~\cite{landau63}:
\ba
S_{cc}=1+2ik_cf_{cc};\hspace{1cm}  S_{nn}=1+2ik_nf_{nn};\hspace{1cm} S_x=2i\sqrt{k_nk_c}f_x
\ea
Substituting these relations in the expressions (3) one immediately obtains:
\ba
\Psi^*_c(r)=\left (\begin{array}{c}i\frac{e^{-ik_cr}-S_{cc}e^{ik_cr}}{2k_cr} \\ -i\frac{S_{nc}e^{ik_nr}}{2\sqrt{k_ck_n}r} \end{array}\right)\hspace{1cm}
\Psi^*_n(r)=\left (\begin{array}{c} -i\frac{S_{nc}e^{ik_cr}}{2r\sqrt{k_ck_n}}\\ i\frac{e^{-ik_nr}-S_{nn}e^{ik_nr}}{2rk_n}  \end{array}\right)
\ea
From the other hand the wave functions $\Psi_c(r)$ and $\Psi_n(r)$ can be constructed as the
linear combination of   two real solutions of  equations (2)
\ba
\Psi^{(1)}=\left(\begin{array}{c}\Psi_c^{(1)}\\\Psi_n^{(1)}\end{array}\right)\hspace{1cm}\Psi^{(2)}=\left (\begin{array}{c}\Psi_c^{(2)}\\\Psi_n^{(2)}\end{array}\right)
\ea
 with the standard  boundary conditions\footnote{In  terms of  the  wave function   $\Phi(r)=\frac{\Psi(r)}{r}$  this condition requires the  regularity   at r=0.}  $ \Psi^{(1)}(0)=\Psi^{(2)}(0)=0.$
Keeping this in mind we will look for  the desired wave functions in the form:
\ba
\Psi^*_c(r)=A_c^{(1)}\Psi^{(1)} +A_c^{(2)}\Psi^{(2)}; \hspace{1cm}
\Psi^*_n(r)=A_n^{(1)}\Psi^{(1)} +A_n^{(2)}\Psi^{(2)}
\ea
where $A_c^{(1)},A_c^{(2)},A_n^{(1)},A_n^{(2)}$ are arbitrary complex numbers.\\
Substituting  the expressions (6),(7)   in  (1) one gets:
\ba
M_c&=&A_c^{(1)}\int \left (\Psi_c^{(1)}M_c^{(0)} +\Psi_n^{(1)}M_n^{(0)}\right )d^3r\nn
&+&A_c^{(2)}\int \left (\Psi_c^{(2)}M_c^{(0)} +\Psi_n^{(2)}M_n^{(0)}\right )d^3r
=A_c^{(1)}I_1+A_c^{(2)}I_2\nn
M_n&=&A_n^{(1)}\int \left (\Psi_c^{(1)}M_c^{(0)} +\Psi_n^{(1)}M_n^{(0)}\right )d^3r\nn
&+&A_n^{(2)}\int \left (\Psi_c^{(2)}M_c^{(0)} +\Psi_n^{(2)}M_n^{(0)}\right )d^3r
=A_n^{(1)}I_1+A_n^{(2)}I_2
\ea
Making use  that any  real  solution  of  equations  (2)  out of the potential range $(U_{ij}=0)$
 can be taken in  the form:
\ba
\Psi(r)=\frac{g\sin(kr+\delta(k))}{kr}=\frac{g}{2ikr}\left (e^{ikr+i\delta(k)}-e^{-ikr-i\delta(k)}\right)
\ea
 we will look for  the real solutions  out of potential range as
\footnote{ We consider only the class of strong potentials with the sharp boundary.}:
\ba
 \Psi_c^{(1)}(r)&=&\frac{g_1e^{ik_cr}-g_1^*e^{-ik_cr}}{2ik_cr};\hspace{1cm}
 \Psi_c^{(2)}(r)=\frac{g_2e^{ik_cr}-g_2^*e^{-ik_cr}}{2ik_cr}\nn
 \Psi_n^{(1)}(r)&=&\frac{h_1e^{ik_nr}-h_1^*e^{-ik_nr}}{2ik_nr};\hspace{1cm}
 \Psi_n^{(2)}(r)=\frac{h_2e^{ik_nr}-h_2^*e^{-ik_nr}}{2ik_nr}
\ea
In order to obtain the relations between the unknown coefficients\footnote{These factors are functions of
pions momenta $k_c,k_n$}   in the above expressions let us at first  compare   the asymptotic  behavior of the
initial wave functions  $\Psi_c(r)$  in (5) with the first raw in the  parametrization  (10):
\ba
\frac{A_c^{(1)}}{2ik_c}\left (g_1e^{ik_cr}-g_1^*e^{-ik_cr}\right )&+&\frac{A_c^{(2)}}{2ik_c}\left (g_2e^{ik_cr}-g_2^*e^{-ik_cr}\right )=
\frac{i}{2k_c}\left (e^{-ik_cr}-S_{cc}e^{ik_cr}\right )\nn
\frac{A_c^{(1)}}{2ik_n}\left (h_1e^{ik_nr}-h_1^*e^{-ik_nr}\right )&+&\frac{A_c^{(2)}}{2ik_n}\left (h_2e^{ik_nr}-h_2^*e^{-ik_nr}\right )=
-\frac{i}{2\sqrt{k_nk_c}}S_{cn}e^{ik_nr}\nn
\ea
Gathering the structures in front of  the appropriate  exponents and solving the system of  obtained equations
after a bit algebra we get:
\ba
A_c^{(1)}&=&\frac{h_2^*}{H};\hspace{1cm} A_c^{(2)}=-\frac{h_1^*}{H};\hspace{1cm}H=g_1^*h_2^*-h_1^*g_2^*;\nn
S_{cc}&=&\frac{h_2^*g_1-h_1^*g_2}{H}; \hspace{1cm} S_{cn}=\sqrt{\frac{k_c}{k_n}}\frac{h_2^*h_1-h_1^*h_2}{H}
\ea
Carry out the same procedure for $\Psi(r)$     we obtain the relevant  relations for  the case of kaon decay to pair of neutral pions:
\ba
A_n^{(1)}&=&-\frac{g_2^*}{H};\hspace{1cm} A_n^{(2)}=-\frac{g_1^*}{H};\nn
S_{nn}&=&-\frac{g_2^*h_1-g_1^*h_2}{H}; \hspace{1cm} S_{nc}=\sqrt{\frac{k_n}{k_c}}\frac{g_1^*g_2-g_1g_2^*}{H}
\ea
In respect that  due to T-invariance   $S_{cn}=S_{nc}$   it can be   checked  that obtained  relations
 satisfied the  unitarity  constraints:
\ba
|S_{nn}|^2+|S_{nc}|^2=1;\hspace{1cm} |S_{cc}|^2+|S_{cn}|^2=1
\ea
As has been seen from expression   (9)  the imaginary parts  of  functions  $g_{1(2)}, h_{1(2)}$
are  determined by appropriate phases.\footnote{The phases are odd functions of relevant momenta
$\delta(-k)=-\delta(k)$ .} For instance,  from  first equation  in (10):
 $$g_1=ge^{i\delta(k_c)}=g\cos{\delta(k_c)}+ig\sin{\delta(k_c)}=
 g\cos{\delta(k_c)}+ik_cg\frac{\sin{\delta(k_c)}}{k_c}$$ At considered low energy  one can safely confined by linear term in phases dependence on momenta :
\ba
g_1&=&d_c^{(1)}+ik_ca_c^{(1)};\hspace{1cm}g_2=d_c^{(2)}+ik_ca_c^{(2)}\nn
h_1&=&d_n^{(1)}+ik_na_n^{(1)};\hspace{1cm}h_2=d_n^{(2)}+ik_na_n^{(2)}
\ea
Substituting these  relations in expressions (12), (13) after cumbersome, but simple algebra we obtain the energy dependence of  S-matrix elements in the two channel case\footnote{The similar expressions  are cited   in the textbook~\cite{baz71},  but with wrong numerator in  the  inelastic case.}:
\ba
S_{cc}&=&\frac{(1+ik_ca_{cc})(1-ik_na_{nn})-k_nk_ca_x^2}
{(1-ik_ca_{cc})(1-ik_na_{nn})+k_nk_ca_x^2}\nn
S_{nn}&=&\frac{(1+ik_na_{nn})(1-ik_ca_{cc})-k_nk_ca_x^2} {(1-ik_ca_{cc})(1-ik_na_{nn})+k_nk_ca_x^2}\nn
S_{cn}&=&S_{nc}=\frac{2i\sqrt{k_ck_n}a_x}{(1-ik_ca_{cc})(1-ik_na_{nn})+k_nk_ca_x^2}
\ea
where :
\ba
a_{nn}&=&\frac{a_n^{(2)}d_c^{(1)}-a_n^{(1)}d_c^{(2)}}{d_n^{(2)}d_c^{(1)}-d_n^{(1)}d_c^{(2)}};
\hspace{1.cm}
a_{cc}=\frac{a_c^{(1)}d_n^{(2)}-a_c^{(2)}d_n^{(1)}}{d_n^{(2)}d_c^{(1)}-d_n^{(1)}d_c^{(2)}};\nn
a_x&=&\frac{a_n^{(1)}d_n^{(2)}-a_n^{(2)}d_n^{(1)}}{d_n^{(2)}d_c^{(1)}-d_n^{(1)}d_c^{(2)}}=
\frac{a_c^{(2)}d_c^{(1)}-a_c^{(1)}d_c^{(2)}}{d_n^{(2)}d_c^{(1)}-d_n^{(1)}d_c^{(2)}}
\ea
Now we are in the position to get  the dependence of matrix elements (1) on pairs momenta $k_c,k_n$.
Introducing   the  real   combinations\footnote{The integrals $I_{1,2}$ are real quantities. }:
\ba
M_{0c}=\frac{I_1d_n^{(2)}-I_2d_n^{(1)}}{d_n^{(2)}d_c^{(1)}- d_n^{(1)}d_c^{(2)}}\hspace{1cm}
M_{0n}=\frac{-I_1d_c^{(2)}+I_2d_c^{(1)}}{d_n^{(2)}d_c^{(1)}- d_n^{(1)}d_c^{(2)}}
\ea
and making use  the expressions   (8),(12), (13),(15)    we obtain our final result :
\ba
M_c&=&M_{0c}\frac{1-ik_na_{nn}}{D}+ik_nM_{0n}\frac{a_x}{D}\hspace{1cm}
M_n=M_{0n}\frac{1-ik_ca_{cc}}{D}+ik_cM_{0c}\frac{a_x}{D}\nn
D&=&(1-ik_ca_{cc})(1-ik_na_{nn})+k_nk_ca_x^2
\ea
For applications it is more convenient to rewritten these relations through the amplitudes of elastic  pion-pion scattering $f_{cc},f_{nn}$ and  charge exchange $f_x$:
\ba
M_c&=&M_{0c}(1+ik_cf_{cc})+ik_nM_{0n}f_x; \hspace{1cm}
M_n=M_{0n}(1+ik_nf_{nn})+ik_cM_{0c}f_x\nn
f_{cc}&=&\frac{a_{cc}(1-ik_na_{nn})+ik_na_x^2}{D};\hspace{0.5cm} f_{nn}=\frac{a_{nn}(1-ik_ca_{cc})+ik_ca_x^2}{D};\hspace{0.5cm}f_x=\frac{a_x}{D};\nn
\ea
These relations  expressing  the decay matrix elements (1)  through the amplitudes of pion-pion scattering
are the main result of present work. Their application to  $K\to 3\pi$ and $K^\pm\to \pi^+\pi^-e^{\pm}\nu$ decays
allow us ~\cite{gevorkyan07,  gevorkyan2} to take  into account the electromagnetic interaction among the charged pions in the final state for any  invariant mass of the pion pair.\\
The first  terms in the expansion  (20) coincide  with appropriate expressions  in ~\cite{cabibbo04,cabibbo05}, i.e. the  $M_{0c}, M_{0n}$  introduced above (see eq.  (18))  can be interpreted  as   so called "unperturbed" amplitudes introduced in ~\cite{cabibbo04}.\\
 The two channel task  considered  in the present  work  permits  to estimate the accuracy of the scattering lengths values extracting from experimental data on kaons decays. Moreover obtained  expressions  allows one to correctly take into account the electromagnetic effects in the final state not only above the charged pions production threshold, but also
 for bound states.~\cite{gevorkyan07,gevorkyan2}
We are grateful to V.D.~Kekelidze  who initiated and support this work  and  D.T.~Madigozhin  for many stimulating and useful discussions.

\end{document}